\documentclass[twocolumn,showpacs,preprintnumbers,amsmath,amssymb,prb]{revtex4}

\usepackage{graphicx}
\usepackage{dcolumn}
\usepackage{bm}

\begin{document}


\title{Order-disorder effect of $A$-site and oxygen-vacancy on magnetic and transport properties of Y$_{1/4}$Sr$_{3/4}$CoO$_{3-\delta}$}

\author{Shun Fukushima}
\author{Tomonori Sato}%
\author{Daisuke Akahoshi}%
\author{Hideki Kuwahara}%
 \email{h-kuwaha@sophia.ac.jp}
\affiliation{%
Department of Physics, Sophia University \\
Chiyoda-ku, Tokyo 102-8554, JAPAN
}%

\date{\today}

\begin{abstract}
We have synthesized $A$-site ordered ($A$O)- and $A$-site disordered ($A$D)-Y$_{1/4}$Sr$_{3/4}$CoO$_{3-\delta}$ (YSCO) with various oxygen deficiency $\delta$, and have made a comparative study of the structural and physical properties. 
We have found that $A$-site (Y/Sr) ordering produces the unconventional oxygen-vacancy ordered (OO) structure, and that the magnetic and transport properties of both $A$O- and $A$D-YSCO strongly depend on the oxygen-vacancy (or excess oxygen) ordering pattern. 
$A$O-YSCO with a stoichiometric $\delta$ of 0.5 has the unconventional OO structure reflecting Y/Sr ordering pattern. 
With decreasing $\delta$ from 0.5, the overall averaged OO structure is essentially unchanged except for an increase of occupancy ratio for the oxygen-vacant sites. 
At $\delta = 0.34$, excess oxygen atoms are ordered to form a novel superstructure, which is significant for the room-temperature ferromagnetism of $A$O-YSCO\@. 
In $A$D-YSCO, on the other hand, the quite different OO structure, which is of a brownmillerite-type, is found only in the vicinity of $\delta = 0.5$. 
\end{abstract}

\maketitle

\section{\label{Introduction}Introduction}

Transition-metal oxides with perovskite structure, $R_{1-x}Ae_xB$O$_3$ ($R$:\ rare-earth, $Ae$:\ alkaline-earth, and $B$:\ transition-metal), and their derivatives exhibit rich physical properties such as high-$T_C$ superconductivity in Cu oxides and colossal magnetoresistance (CMR) effect in Mn oxides.\cite{Imada_RMP_70} 
Perovskite-related oxides have been intensively studied from the viewpoints of not only strongly correlated electron physics but also potential application for correlated electron devices. 
In normal perovskite structures, $R$ and $Ae$ atoms randomly occupy the $A$-sites. 
Resultant $A$-site randomness often suppresses electronic phases, resulting in a large reduction of phase transition temperatures, which makes its practical application difficult. 
Therefore, it is required to reduce or remove such $A$-site randomness for the achievement of its practical application. 

Perovskite oxides with $A$-site ordered ($A$O) structures are expected as promising candidates for correlated electron devices because they are free from $A$-site randomness. 
One of the typical examples is a high-$T_C$ superconductor YBa$_2$Cu$_3$O$_7$, which has a relatively high superconducting transition temperature of 90~K among high-$T_C$ superconductors. 
Another example is $A$O-$R$BaMn$_2$O$_6$, in which the charge- and orbital-ordering temperature ($T_{\rm CO} = 300$-480~K) is much higher than that of conventional manganites with $A$-site disordered ($A$D) structures.\cite{Nakajima_JPCS_63,Nakajima_JPSJ_71} 
$A$-site cation ordering not only raises the $T_{\rm CO}$ but also gives birth to an anomalous electronic phase that is not found in $A$D perovskite manganites. 
In $A$O-$R$BaMn$_2$O$_6$, for example, the charge- and orbital-ordering pattern and the associated magnetic structure are quite different from those of $A$D manganites.\cite{Arima_PRB_66,Williams_PRB_66,Uchida_JPSJ_71,Kageyama_JPSJ_72} 
Furthermore, in $A$O-$R$BaMn$_2$O$_6$, the electronic phases such as the ferromagnetic metallic, charge- and orbital-ordered, and $A$-type antiferromagnetic phases compete with each other to form a multicritical point at $R$ = Nd\@. 
$R$/Ba disordering largely modifies the electronic phase diagram.\cite{Akahoshi_PRL_90,Nakajima_JPSJ_73} 
In $A$D-$R$BaMn$_2$O$_6$, the ferromagnetic metallic phase and charge- and orbital-ordered one are largely suppressed, and consequently, large phase fluctuation is enhanced near the original multicritical region ($R$ = Nd). 
Such large phase fluctuation is significant for the CMR effect.\cite{Akahoshi_PRL_90,Motome_PRL_91} 

A cobaltite with a new type of an $A$O perovskite structure, Y$_{1-x}$Sr$_x$CoO$_{3-\delta}$ has been recently reported by Istomin {\it et al}.\cite{Istomin_Chem.Mater._15} and Withers {\it et al}.\cite{Withers_JSSC_174} 
In Fig.~1(d), the crystal structure of $A$O-Y$_{1-x}$Sr$_x$CoO$_{3-\delta}$ ($x$ = 3/4, $\delta = 0.50$) is displayed. 
Y and Sr atoms are ordered within the $ab$-plane, and a CoO$_6$ octahedral layer and a CoO$_4$ tetrahedral layer alternately stack along the $c$-axis to form four-times periodicity. 
Oxygen-vacancies in the CoO$_4$ tetrahedral layers (black spheres in Figs.~1(d) and  (e)) are regularly arranged in an unconventional way: four oxygen-vacancies form a cluster near Y atoms. 
The oxygen deficiency $\delta$ ($\le 0.5$) depends on occupancy ratio only for the oxygen-vacant sites. 
When the oxygen-vacant sites are fully occupied, $\delta$ becomes zero. 
Kobayashi {\it et al}.\ reported that $A$O-Y$_{1-x}$Sr$_x$CoO$_{3-\delta}$ with $0.75 \le x \le 0.8$ shows a ferromagnetic behavior above room-temperature.\cite{Kobayashi_PRB_72} 
Ishiwata {\it et al}.\ proposed that ordering of Co $e_g$ orbitals causes the room-temperature ferromagnetism.\cite{Ishiwata_PRB_75} 
As another characteristic of $A$O-Y$_{1-x}$Sr$_x$CoO$_{3-\delta}$, the physical properties are susceptible to the variation of $\delta$. 
The ground state of $A$O-Y$_{1/3}$Sr$_{2/3}$CoO$_{3-\delta}$ changes from an antiferromagnetic insulator ($\delta = 0.34$) to a ferromagnetic metal ($\delta = 0.30$) with a slight decrease of $\delta$.\cite{Maignan_JSSC_178} 

In this study, we have prepared $A$O- and $A$D-Y$_{1/4}$Sr$_{3/4}$CoO$_{3-\delta}$ with different $\delta$, and have made a comparative study of their structural and physical properties in order to reveal the effect of $A$-site and oxygen-vacancy arrangement on the physical properties of the perovskite cobaltites.

\section{\label{Experiment}Experiment}

\begin{table}[b]
\caption{Annealing conditions, oxygen deficiency $\delta$ determined by iodometric titration, and corresponding Co valences of Y$_{1/4}$Sr$_{3/4}$CoO$_{3-\delta}$ (YSCO).}
\label{t1}
\begin{ruledtabular}
\begin{tabular}{cdc}
\multicolumn{3}{l}{(a) $A$O/OO-YSCO} \\
\hline
\multicolumn{1}{c}{annealing condition} & \mbox{$\delta$} & \mbox{Co valence} \\
\hline
400 atm O$_2$ 873 K (10 h) & 0.302(17) & 3.15(3) \\
1 atm O$_2$ 773 K (2 h) & 0.340(9) & 3.07(2) \\
Ar 1173 K (12 h) & 0.437(3) & 2.88(1) \\
\hline
\multicolumn{3}{l}{(b) $A$D/OD-YSCO} \\
\hline
\multicolumn{1}{c}{annealing condition} & \mbox{$\delta$} & \mbox{Co valence} \\
\hline
400 atm O$_2$ 873 K (10 h) & 0.152(6) & 3.45(1) \\
1 atm O$_2$ 773 K (2 h) & 0.158(4) & 3.43(1) \\
4\% H$_2$ in Ar 423 K (10 min) & 0.206(7) & 3.34(1) \\
4\% H$_2$ in Ar 473 K (10 min) & 0.222(7) & 3.31(1) \\
4\% H$_2$ in Ar 523 K (10 min) & 0.272(3) & 3.21(1) \\
4\% H$_2$ in Ar 573 K (10 min) & 0.319(5) & 3.11(1) \\
\hline
\multicolumn{3}{l}{(c) $A$D/OO-YSCO} \\
\hline
\multicolumn{1}{c}{annealing condition} & \mbox{$\delta$} & \mbox{Co valence} \\
\hline
4\% H$_2$ in Ar 573 K (6 h) & 0.470(4) & 2.81(1) \\
\end{tabular}
\end{ruledtabular}
\end{table}

$A$O- and $A$D-Y$_{1/4}$Sr$_{3/4}$CoO$_{3-\delta}$ (YSCO) were prepared in a polycrystalline form by solid state reaction. 
Mixed powders of Y$_2$O$_3$, SrCO$_3$, and CoO were heated at 1073~K in air with a few intermediate grindings, and sintered at 1423~K in air. 
The sintered powder was then treated at 1173~K in Ar to enhance $A$-site ordering. 
The resultant product has the $A$O-structure with $\delta = 0.44$. 
$A$D-YSCO was obtained by rapidly cooling the sintered powder from 1473 (air) to 77~K (liquid nitrogen). 
Hereafter, we refer to the thus obtained sample as RC-YSCO\@. 
We controlled $\delta$ of $A$O- and $A$D-YSCO by annealing the samples under various conditions (Table~I)\@. 
The values of $\delta$ were determined by iodometric titration. 
Each powdered sample (ca.~30 mg) was dissolved in 1~M HCl solution (ca.~50 ml) containing excess aqueous KI\@. 
The amount of the formed I$_2$ was titrated with 0.01~M Na$_2$S$_2$O$_3$ solution using starch as a colorimetric indicator. 
The titration was performed more than three times for each sample. 
Co valences  were determined by assuming that the samples have a nominal cation ratio. 
The average values of $\delta$ and the corresponding Co valences are listed in Table~I\@. 
Powder X-ray diffraction (XRD) measurements were carried out at room-temperature using a RIGAKU Rint-2100 diffractometer with Cu $K\alpha$ radiation. 
The crystal structures of the samples were analyzed by Rietveld method using RIETAN-2000.\cite{rietan} 
Magnetic and transport properties were measured from 5 to 350~K using a Quantum Design, Physical Property Measurement System (PPMS)\@. 
Magnetization measurements above room-temperature ($T = 300$-800~K) were carried out using a Quantum Design, Magnetic Property Measurement System (MPMS)\@.

\section{\label{Results}Results}

\subsection{\label{Crystal structure}Crystal structures}

\begin{figure*}[tb]
\begin{center}
\includegraphics[scale=1.00,clip]{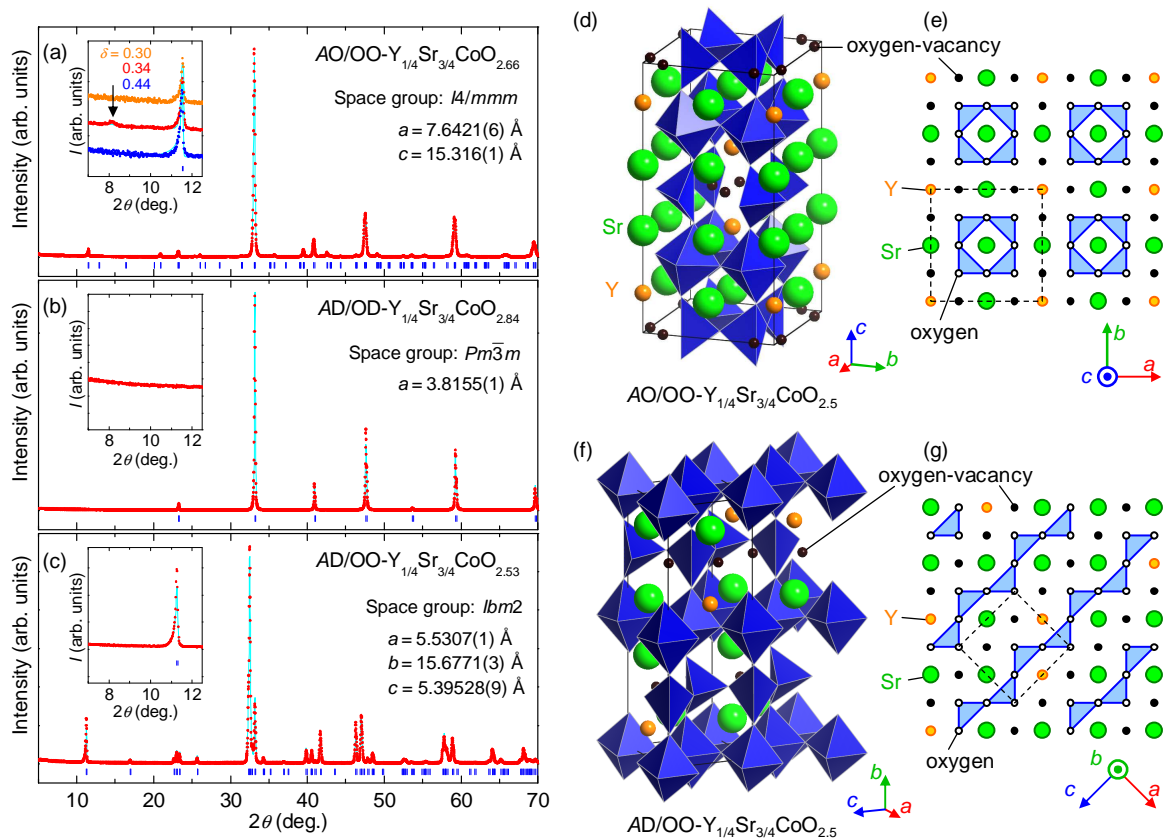}
\end{center}
\caption{(Color online) X-ray-diffraction patterns of (a) $A$-site and oxygen-vacancy ordered ($A$O/OO)-Y$_{1/4}$Sr$_{3/4}$CoO$_{2.66}$, (b) $A$-site and oxygen-vacancy disordered ($A$D/OD)-Y$_{1/4}$Sr$_{3/4}$CoO$_{2.84}$, and (c) $A$-site disordered and oxygen-vacancy ordered ($A$D/OO)-Y$_{1/4}$Sr$_{3/4}$CoO$_{2.53}$. Solid circles and solid lines represent observed and calculated diffraction profiles, respectively. Vertical marks under the diffraction profiles indicate calculated peak positions. The insets show the diffraction profiles at low angles. For comparison, diffraction profiles of $\delta = 0.30$ and 0.44 ($A$O/OO) are also depicted in the inset of (a). Crystal structure and its idealized oxygen-vacant layer of (d),(e) $A$O/OO-Y$_{1/4}$Sr$_{3/4}$CoO$_{2.5}$ and (f),(g) $A$D/OO-Y$_{1/4}$Sr$_{3/4}$CoO$_{2.5}$. Small black spheres of (d)-(g) represent oxygen-vacancies.}
\label{f1}
\end{figure*}

Figure 1(a) shows the XRD pattern of $A$O-YSCO with $\delta = 0.34$. 
It should be noted that the Bragg peak around $2\theta = 11.5$ deg.\ is a piece of evidence that both Y/Sr and oxygen-vacancies are ordered. 
Except for an additional Bragg peak indicated by an arrow in the inset of Fig.~1(a), the XRD patterns of $A$O-YSCO with $\delta = 0.30$ and 0.44 are similar to that of $\delta = 0.34$ ($A$O/OO), and all the XRD patterns can be fitted to the $A$O/OO structure ($I4/mmm$) shown in Fig.~1(d). 
Note that all $A$O/OO-YSCO prepared in this study, whose $\delta$ is smaller than the stoichiometric composition ($\delta = 0.5$), have excess oxygen atoms. 
The result of the XRD measurement indicates that these excess oxygen atoms randomly occupy the oxygen-vacant sites (or raise the occupancy ratio of the site) in the CoO$_4$ layers with keeping the overall averaged OO structure shown in Figs.~1(d) and (e), that is, the OO structure is robust against the variation of $\delta$. 
The additional Bragg peak observed in $\delta = 0.34$ ($A$O/OO) can be indexed as (1/4 1/4 0) in an $a_p \times a_p \times a_p$ setting ($a_p$ denotes a pseudo-cubic perovskite cell), evidencing the existence of a four-times superstructure along the [1 1 0] direction. 
Because the additional peak is sensitive to $\delta$ and found only in the XRD pattern of $\delta = 0.34$ ($A$O/OO), it is reasonable to conclude that the superstructure arises from ordering of the excess oxygen atoms occupying the oxygen-vacant sites. 
This means that $A$O-YSCO probably has another oxygen stoichiometry near $\delta = 0.34$ besides the stoichiometric $\delta = 0.5$. 
$\delta = 1/3$ (or $3/8$), which is close to 0.34, is likely to be the second stoichiometric composition. 
A similar superlattice peak (1/4 1/4 0) is also observed in $A$O-Er$_{0.78}$Sr$_{0.22}$CoO$_{2.63}$ by use of a high-intensity synchrotron X-ray source \cite{Ishiwata_PRB_75}, implying that $\delta = 0.34$ ($A$O/OO) has the similar oxygen superstructure to $A$O-Er$_{0.78}$Sr$_{0.22}$CoO$_{2.63}$. 
The intensity of the superlattice peak of our sample is much stronger than that of $A$O-Er$_{0.78}$Sr$_{0.22}$CoO$_{2.63}$\cite{fnote_ESCO}, indicating that the oxygen deficiency $\delta$ of our sample is closer to the stoichiometry than that of $A$O-Er$_{0.78}$Sr$_{0.22}$CoO$_{2.63}$. 
The detailed crystal structure of $\delta = 0.34$ ($A$O/OO) is now under investigation. 

Then, we exhibit the XRD pattern of RC-YSCO with $\delta = 0.16$ in Fig.~1(b). 
What is a significant difference from the XRD patterns of $A$O-YSCO system is that the characteristic Bragg peaks arising from $A$-site and oxygen-vacancy ordering totally vanish. 
The XRD pattern can be well fitted to a simple cubic perovskite structure ($Pm\bar{3}m$), indicating that Y and Sr atoms randomly occupy the $A$-sites, and that oxygen-vacancies are randomly distributed in the structure; RC-YSCO has an $A$D and oxygen-vacancy disordered (OD) structure. 
All RC-YSCO with $0.15 \le \delta \le 0.32$ have the same simple cubic structure as RC-YSCO with $\delta = 0.16$. 
From now on, we will refer to these compounds as $A$D/OD-YSCO\@. 

On the other hand, the XRD pattern of RC-YSCO with $\delta$ = 0.47 (Fig.~1(c)) is quite different from those of RC-YSCO with $0.15 \le \delta \le 0.32$. 
A characteristic Bragg peak, which implies $A$-site and/or oxygen-vacancy ordering, is clearly seen around $2\theta = 11.5$ deg. 
By annealing RC-YSCO with $\delta = 0.47$ in O$_2$ at 573~K, which is low enough to prevent $A$-site cations from moving around to rearrange, the characteristic peak totally disappears, and the annealed compound takes a simple cubic perovskite structure. 
Therefore, we conclude that only oxygen-vacancy ordering contributes to the appearance of the characteristic peak; RC-YSCO with $\delta = 0.47$ has an $A$D/OO structure. 
Rietveld analysis reveals that the OO structure of $A$D-YSCO with $\delta = 0.47$ is of a brownmillerite (Ca$_2$(Fe,Al)$_2$O$_5$)-type (Fig.~1(f) and (g), $Ibm2$)\cite{Colville_Acta_Cryst_B27}, which is quite different from that of $A$O/OO-YSCO shown in Figs.~1(d) and (e). 

To summarize this section, YSCO can be classified into three types of the crystal structures: the $A$O/OO, $A$D/OD, and $A$D/OO structures. 
In the following section, we demonstrate that the arrangement of Y/Sr and oxygen-vacancies considerably affects the physical properties of YSCO\@.

\subsection{\label{AOOO}$A$-site and oxygen-vacancy ordered ($A$O/OO)-YSCO}

\begin{figure}[tb]
\begin{center}
\includegraphics[scale=0.50,clip]{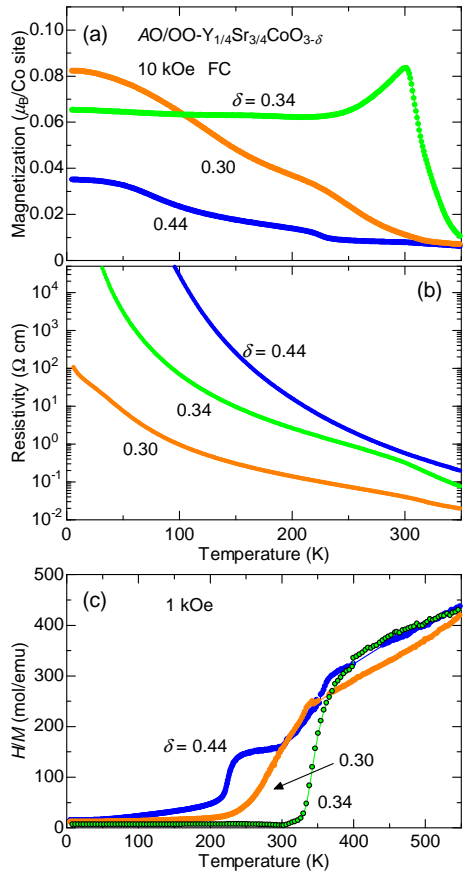}
\end{center}
\caption{(Color online) Temperature dependence of (a) magnetization, (b) resistivity, and (c) inverse magnetization ($H/M$) of $A$O/OO-Y$_{1/4}$Sr$_{3/4}$CoO$_{3-\delta}$. FC:\ field cooled.}
\label{f2}
\end{figure}

Figures 2(a) and (b) show the temperature dependence of the magnetization and resistivity of $A$O/OO-YSCO with $\delta = 0.30$, 0.34, and 0.44. 
The magnetization of $\delta$ = 0.34 ($A$O/OO) shows a weak ferromagnetic transition around 330~K as previously reported by Kobayashi {\it et al}.\cite{Kobayashi_PRB_72} 
Ishiwata {\it et al}.\ suggest that the weak ferromagnetism (the room-temperature ferromagnetism) originates in canted antiferromagnetic or ferrimagnetic order caused by ordering of Co $e_g$ orbital.\cite{Ishiwata_PRB_75} 
Then, the magnetization of $\delta = 0.34$ ($A$O/OO) shows a sharp cusp around 300~K, implying that a magnetic or spin-state transition occurs. 
The resistivity of $\delta = 0.34$ ($A$O/OO) exhibits an anomaly around the weak ferromagnetic transition temperature, below which it is insulating. 
Figure 2(c) shows the inverse magnetization of $A$O/OO-YSCO with $\delta = 0.30$, 0.34, and 0.44. 
All the inverse magnetizations approximately obey the Curie-Weiss law above 400~K\@. 
The Weiss temperature $\theta_W$ and effective moment $P_{\rm eff}$ of $\delta = 0.34$ ($A$O/OO) are found to be $-60$~K and 3.3 $\mu_{\rm B}/{\rm Co}$, respectively. 
The magnetic and transport properties of $\delta = 0.34$ ($A$O/OO) observed in this study are consistent with those previously reported by Kobayashi {\it et al}.\cite{Kobayashi_PRB_72,Kobayashi_JPSJ_75}, except for the magnetic cusp at 300~K\@. 
The origin of the cusp will be discussed later.

\begin{figure}[tb]
\begin{center}
\includegraphics[scale=0.50,clip]{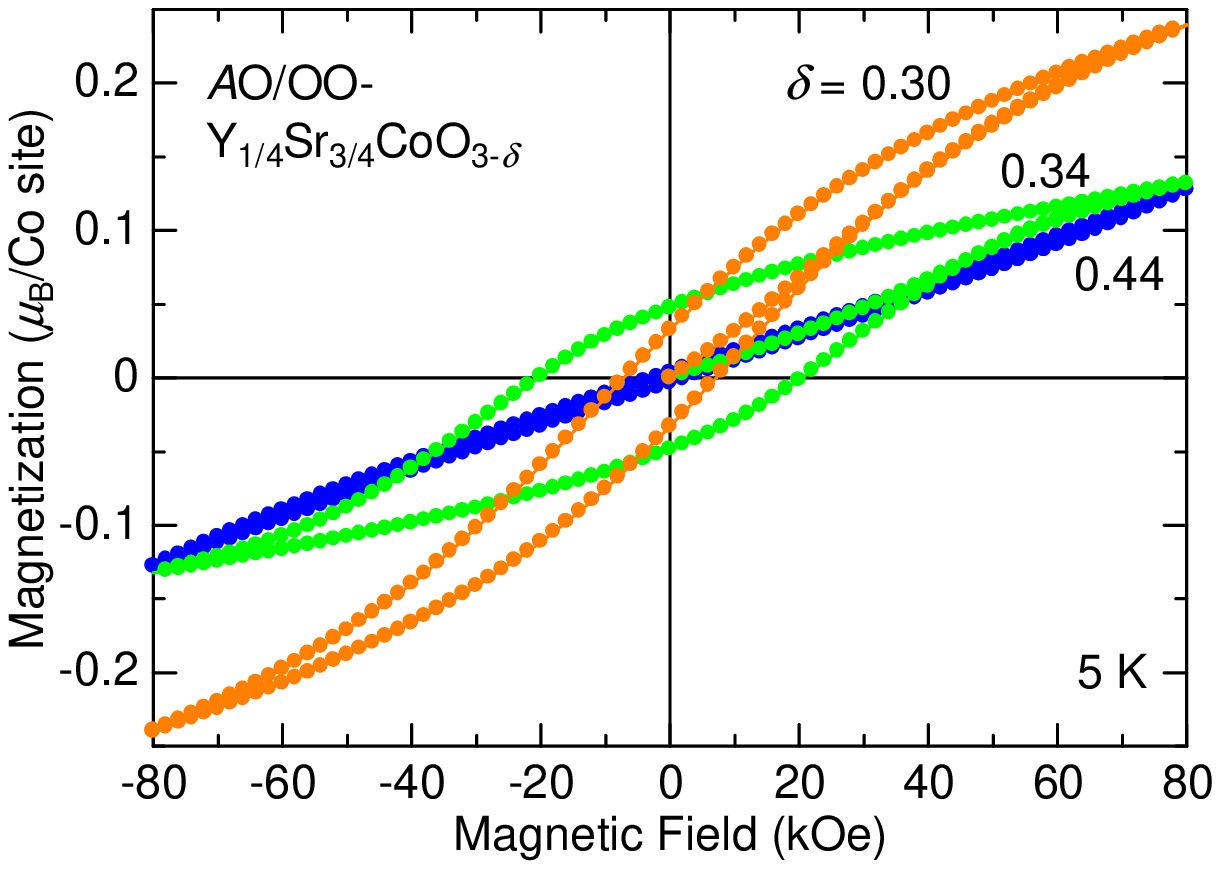}
\end{center}
\caption{(Color online) Magnetic field dependence of magnetization ($M$-$H$ curves) of $A$O/OO-Y$_{1/4}$Sr$_{3/4}$CoO$_{3-\delta}$ at 5 K\@.}
\label{f3}
\end{figure}

In $\delta = 0.30$ ($A$O/OO), the weak ferromagnetic insulating phase above room-temperature is considerably suppressed, while the magnetization is continuously increasing with decreasing temperature, and the ferromagnetic correlation is larger than that of $\delta = 0.34$ ($A$O/OO) below $\sim 100$~K (Figs.~2(a) and 3). 
With decreasing $\delta$ from 0.34 to 0.30, the $\theta_W$ increases from $-60$~K to 60~K, that is, an antiferromagnetic interaction between Co ions turns into a ferromagnetic one. 
The resistivity of $\delta = 0.30$ ($A$O/OO) largely drops in the whole temperature region compared with that of $\delta = 0.34$ ($A$O/OO)\@. 
These results show that $A$O-YSCO approaches a ferromagnetic metal with a decrease of $\delta$, i.e.\ with an increase of Co valence, and that the weak ferromagnetic insulating phase ($\delta \approx 0.34$) and ferromagnetic metallic clusters (oxygen rich region: $\delta \ll 0.34$) coexist in $\delta = 0.30$ ($A$O/OO)\@.  

On the other hand, the magnetization of $\delta = 0.44$ ($A$O/OO), which is close to the stoichiometric composition of $\delta = 0.5$ and has excess oxygen atoms occupying 12~\% of the oxygen-vacant sites, shows a slight increase around 230~K, suggesting that a weak ferromagnetic transition occurs. 
However, as seen from the magnetic field dependence of the magnetization ($M$-$H$ curves) (Fig.~3), the weak ferromagnetic magnetization is much smaller than that of $\delta = 0.34$ ($A$O/OO)\@. 
The inverse magnetization of $\delta = 0.44$ ($A$O/OO) shows a clear anomaly around 350~K (Fig.~2(c)), which is attributed to the remnant of the room-temperature ferromagnetic phase most stabilized around $\delta = 0.34$. 
These results indicate that the magnetic properties of $\delta = 0.44$ ($A$O/OO) can be explained by the coexistence of the matrix phase ($\delta = 0.5$) and embedded clusters ($\delta \approx 0.34$) due to the excess oxygen atoms. 
The weak ferromagnetism below 230~K probably comes from the matrix phase. 

To summarize this section, the room-temperature ferromagnetism is observed only in $\delta = 0.34$ ($A$O/OO), and a slight deviation of $\delta$ from 0.34 strongly suppresses the room-temperature ferromagnetism. 
The sensitivity of the room-temperature ferromagnetic phase to $\delta$ also supports our aforementioned conclusion that $A$O/OO-YSCO has another oxygen stoichiometry around $\delta = 0.34$.

\subsection{\label{ADOD}$A$-site and oxygen-vacancy disordered ($A$D/OD)-YSCO}

\begin{figure}[tb]
\begin{center}
\includegraphics[scale=0.50,clip]{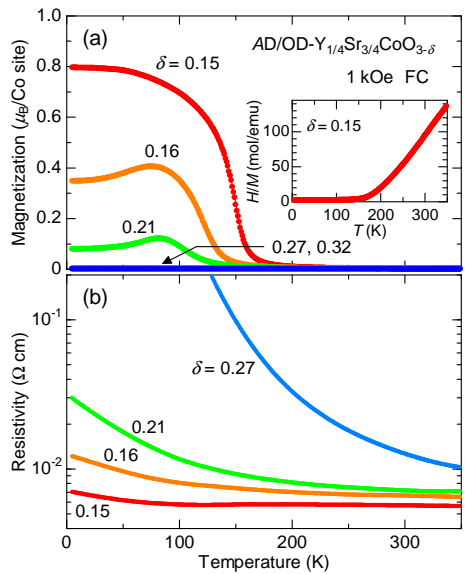}
\end{center}
\caption{(Color online) Temperature dependence of (a) magnetization and (b) resistivity of $A$D/OD-Y$_{1/4}$Sr$_{3/4}$CoO$_{3-\delta}$. The inset of (a) shows the inverse magnetization of $A$D/OD-Y$_{1/4}$Sr$_{3/4}$CoO$_{3-\delta}$ ($\delta = 0.15$)}
\label{f4}
\end{figure}

We display in Figs.~4(a) and (b) the temperature dependence of the magnetization and resistivity of $A$D/OD-YSCO with $0.15 \le \delta \le 0.32$. 
In $\delta = 0.15$ ($A$D/OD), the resistivity exhibits metallic transport, except for a slight upturn at low temperatures, and the magnetization abruptly increases below 160~K\@. 
The $\theta_W$ ($= 190$~K) estimated from the inverse magnetization (the inset of Fig.~4(a)) is close to the magnetic transition temperature. 
The $M$-$H$ curve at 5~K exhibits a ferromagnetic behavior with a saturation magnetization of 1.2 $\mu_{\rm B}/{\rm Co}$ (Fig.~5). 
These facts indicate that $\delta = 0.15$ ($A$D/OD) is a typical ferromagnetic metal. 
Such a ferromagnetic metallic behavior is sometimes observed in conventional perovskite cobaltites with $A$D structure such as La$_{1-x}$Sr$_x$CoO$_3$ ($x \ge 0.18$).\cite{Itoh_JPSJ_63,Senaris_JSSC_118} 
It is widely accepted that the origin of the ferromagnetic metallicity of La$_{1-x}$Sr$_x$CoO$_3$ can be explained by the double exchange interaction.\cite{Itoh_JPSJ_63,Kriener_PRB_69,Saitoh_PRB_56,Yamaguchi_JPSJ_64} 
The saturation magnetization of $\delta = 0.15$ ($A$D/OD) is close to that of La$_{0.8}$Sr$_{0.2}$CoO$_3$ ($\sim 1.3$ $\mu_{\rm B}/{\rm Co}$ at 5~K), indicating that Co$^{3+}$ and Co$^{4+}$ of $\delta = 0.15$ ($A$D/OD) in the ferromagnetic metallic state have similar electronic configurations to those of La$_{1-x}$Sr$_x$CoO$_3$.\cite{Senaris_JSSC_118,Kriener_PRB_69} 
Thus, it is reasonable to conclude that the ferromagnetic metallic state of $A$D-YSCO with $\delta = 0.15$ is stabilized via the double-exchange mechanism.

\begin{figure}[tb]
\begin{center}
\includegraphics[scale=0.50,clip]{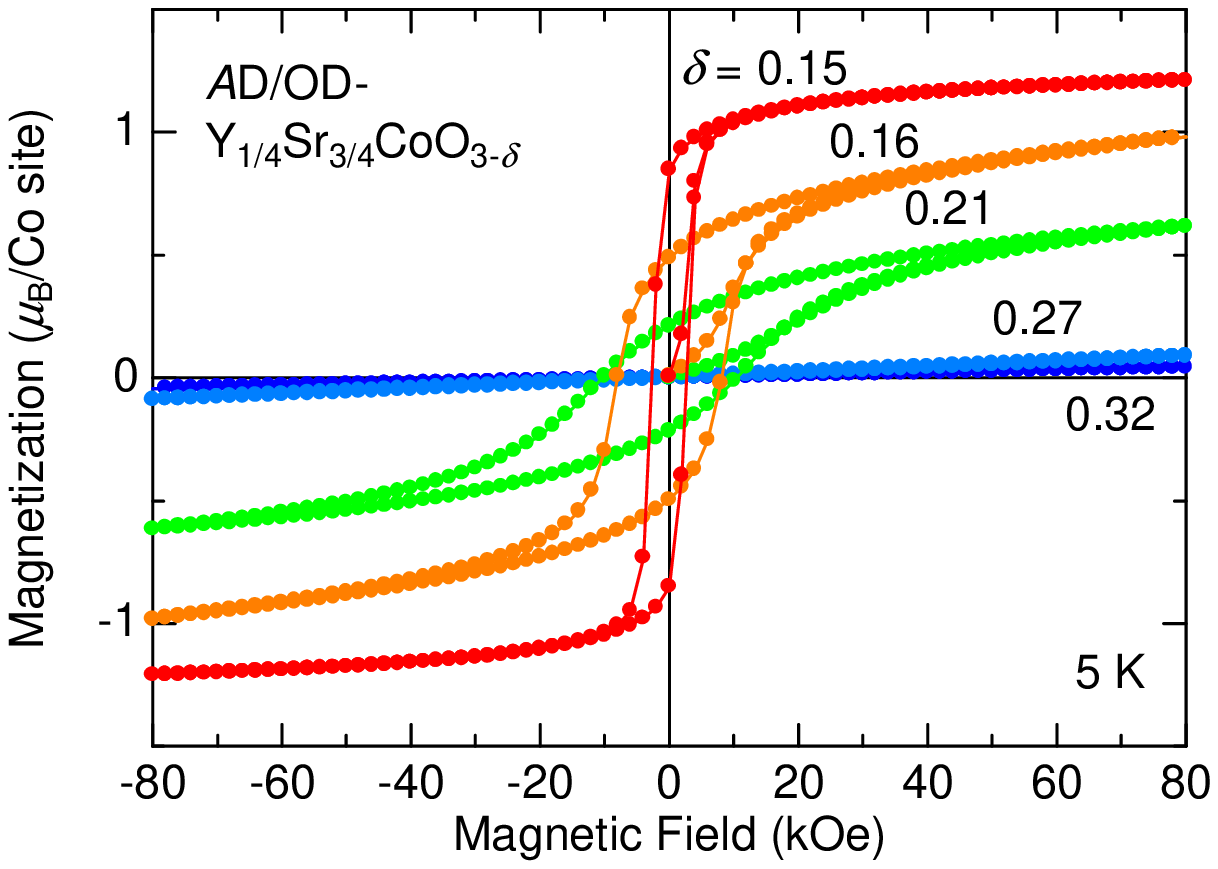}
\end{center}
\caption{(Color online) Magnetic field dependence of magnetization ($M$-$H$ curves) of $A$D/OD-Y$_{1/4}$Sr$_{3/4}$CoO$_{3-\delta}$ at 5 K\@.}
\label{f5}
\end{figure}

With an increase of $\delta$, the ferromagnetic metallic state is steeply suppressed (Fig.~4). 
In $\delta = 0.27$ and 0.32 ($A$D/OD), the ferromagnetic component is negligible as seen from the temperature dependence of the magnetization and the $M$-$H$ curves, and the resistivity exhibits a typical insulating behavior (Figs.~4 and 5).

\subsection{\label{ADOO}$A$-site disordered and oxygen-vacancy ordered ($A$D/OO)-YSCO}

Figure~6(a) shows the temperature dependence of the magnetization and resistivity of $A$D/OO-YSCO with $\delta = 0.47$, which has the brownmillerite-type OO structure as described in section \ref{Crystal structure}\@. 
The magnetization of $A$D/OO-YSCO with $\delta = 0.47$ abruptly increases around 130~K just like that of $A$D/OD-YSCO with $\delta = 0.15$ (Fig.~4(a)). 
However, the saturation magnetization of $0.15$ $\mu_{\rm B}/{\rm Co}$ (Fig.~6(b)) is much smaller than that of $A$D/OD-YSCO with $\delta = 0.15$, and the resistivity of $A$D/OO-YSCO with $\delta = 0.47$ shows an insulating behavior in the whole temperature region. 
The parent compound of $A$D/OO-YSCO, Sr$_2$Co$_2$O$_5$, which also has the brownmillerite-type OO structure, undergoes a similar weak ferromagnetic transition at 200~K and a $G$-type antiferromagnetic transition at 537~K.\cite{Munoz_PRB_78} 
In $A$D/OO-YSCO with $\delta = 0.47$, similarly, a $G$-type anitiferromagnetic transition might occur far above room-temperature.\cite{fnote_HiTemp} 
We note that the weak ferromagnetic moment of $A$D/OO-YSCO with $\delta = 0.47$ is more than 10 times larger than that of Sr$_2$Co$_2$O$_5$. 
In $A$D/OO-YSCO with $\delta = 0.47$, which is in a mixed valence state of Co$^{3+}$/Co$^{2+}$, a ferrimagnetic transition accompanying charge-order might occur at 130~K\@.

\begin{figure}[tb]
\begin{center}
\includegraphics[scale=0.50,clip]{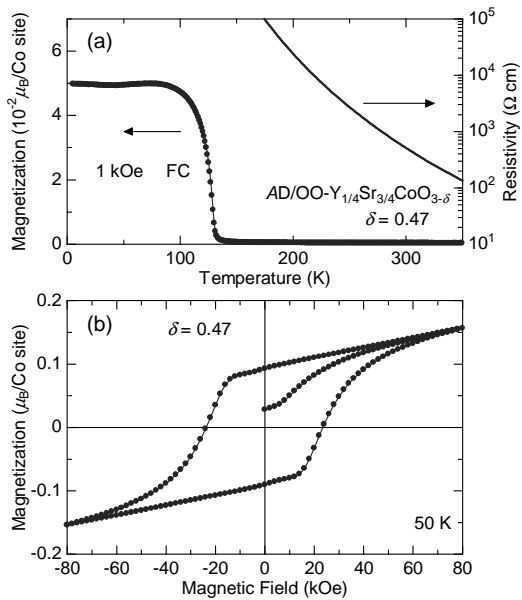}
\end{center}
\caption{(a) Temperature dependence of magnetization and resistivity, and (b) magnetic field dependence of magnetization ($M$-$H$ curve) of $A$D/OO-Y$_{1/4}$Sr$_{3/4}$CoO$_{3-\delta}$ ($\delta = 0.47$).}
\label{f6}
\end{figure}

\section{\label{Discussion}Discussion}

Now we discuss the effect of the $A$-site and oxygen-vacancy arrangement on the physical properties of YSCO. 
$A$O-YSCO with the stoichiometric composition of $\delta = 0.5$ has the unconventional OO structure, in which, four oxygen-vacancies in the oxygen deficient CoO$_{2-2\delta}$ layers form a cluster near Y atom in the adjacent Y$_{1/4}$Sr$_{3/4}$O layers as shown in Figs.~1(d) and (e). 
Y/Sr disordering largely modifies the arrangement of the oxygen-vacancies. 
$A$D-YSCO with $\delta = 0.5$ has the brownmillerite-type OO structure (Figs.~1(f) and (g)), which is often found in conventional perovskite oxides with large oxygen deficiency such as Sr$_2$Fe$_2$O$_5$\cite{Harder_BM}, Ca$_2$Fe$_2$O$_5$\cite{Colville_Acta_Cryst_B26}, and Sr$_2$Co$_2$O$_5$\cite{Takeda_JPSJ_33}. 
From these results, it is obvious that Y/Sr ordering gives rise to the unconventional OO structure reflecting the Y/Sr ordering pattern. 
The OO structure of $A$O-YSCO is so robust that excess oxygen atoms partially occupy the oxygen-vacant sites with keeping the overall OO structure shown in Figs.~1(d) and (e). 
In contrast, the OO structure of $A$D-YSCO (Figs.~1(f) and (g)) is so fragile that it is easily destroyed by introduction of a small amount of excess oxygen atoms. 
Consequently, the OO structure is found only in the vicinity of $\delta = 0.5$ in the case of  $A$D-YSCO. 
The robustness of the OO structure of $A$O-YSCO originates in periodic potential due to $A$-site ordering. 

As for the magnetic properties, both $A$O/OO- and $A$D/OO-YSCO with the stoichiometric $\delta$ (around 0.34 and 0.47 respectively) exhibit the weak ferromagnetic behaviors as demonstrated in section \ref{Results}\@. 
It should be noted that weak ferromagnetic behaviors are also observed in other oxygen stoichiometric $A$O perovskites such as YBaCo$_2$O$_{5+x}$ ($x = 0.50$ and 0.44) and YBaMn$_2$O$_{5+x}$ ($x = 0.50$), which have the superstructures formed by excess oxygen atoms.\cite{Akahoshi_JSSC_156,Karppinen_JSSC_177}\@ 
Therefore, it is probable that oxygen-vacancy (or excess oxygen) ordering plays a significant role in the weak ferromagnetism of both $A$O/OO- and $A$D/OO-YSCO. 

Then, we propose one plausible model to explain the origin of the weak ferromagnetism. 
In the following discussion, we suppose that the magnetic interaction between Co ions is antiferromagnetic. 
Considering a CoO$_6$ octahedron adjacent to a CoO$_4$ tetrahedron (or a CoO$_5$ square pyramid) in oxygen deficient perovskite structures, an inversion symmetry is absent at the center between the two Co ions. 
In such a case, a spin-canted moment is locally induced through the Dzyaloshinskii-Moriya interaction.\cite{Dzyaloshinskii_JPCS_4,Moriya_PhysRev_120} 
Another possible case is as follows: if the spin states of CoO$_6$ and CoO$_4$ (or CoO$_5$) were different from each other, the CoO$_6$-CoO$_4$ (or CoO$_5$) pair could have a local ferrimagnetic moment. 
In the OO structures, the CoO$_6$-CoO$_4$ (or CoO$_5$) pairs, i.e., the local net moments are regularly arranged. 
As a result, the weak ferromagnetic magnetization is macroscopically observed. 
On the other hand, in the OO structure with large oxygen nonstoichiometry (excess oxygen randomly occupy the oxygen-vacant sites) or in the OD structures ($\delta = 0.27$ and 0.32 ($A$D/OD)) in which, the CoO$_6$-CoO$_4$ (or CoO$_5$) pairs are randomly distributed, the local net moments oriented randomly are canceled out with each other. 
The reason why the weak ferromagnetic moment of $\delta = 0.44$ ($A$O/OO) below 230~K is very small may be attributed to randomness due to excess oxygen atoms occupying 12~\% of the oxygen-vacant sites. 
Note that the ferromagnetism of $A$D/OD-YSCO with $\delta = 0.15$ is induced through the double-exchange interaction as described in section \ref{ADOD}\@. 

The magnetization of $\delta = 0.34$ ($A$O/OO) we prepared exhibits the sharp cusp around 300~K (Fig.~2(a)), which is not clearly discerned in $A$O-YSCO previously reported.\cite{Kobayashi_PRB_72} 
In this study, $A$O-YSCO was treated in Ar to enhance Y/Sr ordering as mentioned in \ref{Experiment}, while $A$O-YSCO in the previous reports was not. 
As a result, the degree of Y/Sr order of our sample is higher than that of $A$O-YSCO in the previous reports, which may be the main cause of the emergence of the cusp. 
Another plausible explanation for the emergence of the cusp is that our sample is closer to the oxygen stoichiometry than $A$O-YSCO in the previous reports.

\section{\label{Summary}Summary}

We have synthesized $A$-site ordered ($A$O)- and $A$-site disordered ($A$D)-Y$_{1/4}$Sr$_{3/4}$CoO$_{3-\delta}$ (YSCO) with various oxygen deficiency $\delta$, and have made a comparative study of their structural and physical properties. 
$A$O-YSCO with $\delta = 0.44$, which is near the stoichiometry of $\delta = 0.5$, has the unconventional oxygen-vacancy ordered (OO) structure reflecting $A$-site ordering pattern, and shows the weak ferromagnetic behavior below 230~K\@. 
Decreasing $\delta$ from 0.5 inducing an increase in the occupancy ratio of the oxygen-vacant site does not essentially change the overall OO structure. 
In $A$O-YSCO with $\delta = 0.34$, which is close to another oxygen stoichiometry, excess oxygen ordering in the oxygen-vacant sites causes the room-temperature ferromagnetism. 
On the other hand, in $A$D-YSCO, the different OO structure, which is of the brownmillerite-type, is found only in the vicinity of $\delta = 0.5$, and also shows the weak ferromagnetic behavior below 130~K\@. 
$A$D/OD-YSCO with $\delta = 0.15$ exhibits a ferromagnetic metallic behavior which is attributed to the double-exchange interaction. 
With an increase of $\delta$, that is, with a decrease of Co valence, the ferromagnetic metallic phase is suppressed. 
Our main conclusions are as follows: First, $A$-site ordering gives rise to the unconventional OO structure, and makes the OO structure robust against randomness due to excess oxygen atoms. 
Second, oxygen-vacancy (or excess oxygen) ordering is indispensable for the weak ferromagnetic behaviors of both $A$O- and $A$D-YSCO.

\begin{acknowledgments}
We thank D. Vieweg and A. Loidl for discussions and their help with the magnetization measurements using MPMS\@. 
This work was partially supported by Iketani Science and Technology Foundation, the Mazda Foundation, the Asahi Glass Foundation, and by Grant-in-Aid for Scientific Research (C) from Japan Society for the Promotion of Science (JSPS). 
\end{acknowledgments}



\begin{references}
\bibitem{fnote_HiTemp} Since $A$D/OO-YSCO is chemically unstable, we could not obtain reliable data of the physical properties at higher temperatures.
\end{references}
\end{document}